**Temperature-independent non-linear terahertz transmission by liquid water**

Célia Millon[1], Johannes Schmidt[2], Sashary Ramos[3], Eliane P. van Dam[3], Adrian Buchmann[3], Clara Saraceno[1], Fabio Novelli[3,*]

[1]Photonics and Ultrafast Laser Science, Ruhr University Bochum, 44801 Bochum, Germany

[2]Elettra Sincrotrone Trieste S.C.p.A., 34127, Basovizza, Trieste, Italy

[3]Department of Physical Chemistry II, Ruhr University Bochum, 44801 Bochum, Germany

[*]fabio.novelli@rub.de

*Liquid water is one of the most studied substances, yet many of its properties are difficult to rationalize. The uniqueness of water is rooted in the dynamic network of hydrogen-bonded molecules with relaxation time constants of about one picosecond. Terahertz fields oscillate on a picosecond timescale and are inherently suited to study water. Recent advances in non-linear terahertz spectroscopy have revealed large signals from water, which have been interpreted with different, sometimes competing, theoretical models. Here we show that the non-linear transmission of liquid water at ∼1 THz is equal at 21 °C and 4 °C, thus suggesting that the most appropriate microscopic models should depend weakly on temperature. Among the different mechanisms proposed to date, the resonant reorientation of hydrogen-bonded water molecules might be the most appropriate to describe all of the currently available experimental results.*

On the microscopic scale, water molecules in the liquid phase make hydrogen bonds with tetrahedral structures that fluctuate on the picosecond (ps) timescale[1]. On the macroscopic scale, thermo-dynamic properties of liquid water like density and compressibility are anomalous, as they scale non-continuously[2]. However, it is unclear how the macroscopic behavior emerges from the microscopic properties[3]. An inherently powerful spectroscopic tool to study liquid water is terahertz (THz) radiation because it can reveal the fluctuations of the water network on the picosecond time scale, which are of the same order of magnitude of the relaxation time constants[4]. In fact, THz radiation between about 0.01 and 20 THz is strongly absorbed by the intermolecular, collective modes of liquid water[5–11].

Recently, several non-linear optical techniques in the THz range were applied to the study of liquid water. Optical-pump THz-probe experiments revealed the coupling between a dye and the solvating water[12], the inhomogeneity in the bulk liquid[13,14], and proton quantum effects[15]. THz-Kerr experiments employ a THz pulse to reorient molecules and an optical probe pulse to detect the induced birefringence at visible or near-infrared frequencies, which is modulated by the Raman-active modes[16]. These works revealed the anisotropic polarizability of hydrogen-bonded water molecules[17], the coupling between rotational and translational modes[18], and the effect of salts on the water structure[19]. THz-pump THz-probe experiments are complementary to the aforementioned approaches, because both the pump and the probe interact resonantly with infrared-active modes. While these latter, non-linear THz spectroscopy experiments reported signals with similar sizes at ∼1 THz[20–24] and 12.3 THz[25,26], i.e., a third-order nonlinear response with a magnitude of about $\chi^3 \sim 10^{-13}$ cm$^2$/V$^2$ (see Table 1 in ref.[26]), the results have been rationalized in different ways.

Previously, we performed THz-pump THz-probe experiments at ~12 THz[25,26] and z-scan measurements at ~1 THz on water[24]. In ref.[25], with the aid of molecular dynamics calculations and a dedicated model that includes all possible non-linear effects and every perturbation order, we proposed that the non-linear response of water in the THz range is due to the resonant reorientation of hydrogen-bonded water molecules. The Elsaesser group[22,23] performed THz-pump THz-probe experiments at ~1 THz. They suggested that the THz pump is responsible for the irreversible ionization of water molecules and the subsequent generation of solvated electrons. The Kozlov group[20,21] performed z-scan measurements at ~1 THz on water. Even though liquid water absorbs THz radiation[5–11] and resonant non-linear phenomena are typically stronger than non-resonant ones[27], these authors[20,21] suggested that the non-linear refraction of water at ~1 THz should originate from the non-resonant contribution by oscillations in the mid-infrared, at about 100 THz, 3 microns, or 3300 cm$^{-1}$. At these mid-infrared frequencies, liquid water absorbs radiation via the intramolecular O-H stretch mode[28], rather than through the intermolecular modes involving hydrogen-bonded water molecules[5–11]. The non-resonant model proposed in ref.[20,21] implies that the non-linear and frequency-dependent refraction is proportional to the thermal expansivity squared. For example, see eq.5 in ref.[20] and Figure 4 in ref.[21]. Thus, considering that the expansivity of water depends on temperature and becomes null at[29] 4 °C, one expects that the non-linear refraction by water at ~1 THz should decrease upon cooling.

Here we show that the non-linear transmission of intense radiation at ~1 THz by water is the same at 21 °C and 4 °C, within the experimental uncertainty. This implies that the non-linear absorption coefficient of water is equal at these two temperatures. These results might be of help in understanding which molecular model should be preferred in the description of the non-linear response by liquid water in the THz range, and suggest that models that dependent weakly on temperature should be preferred.

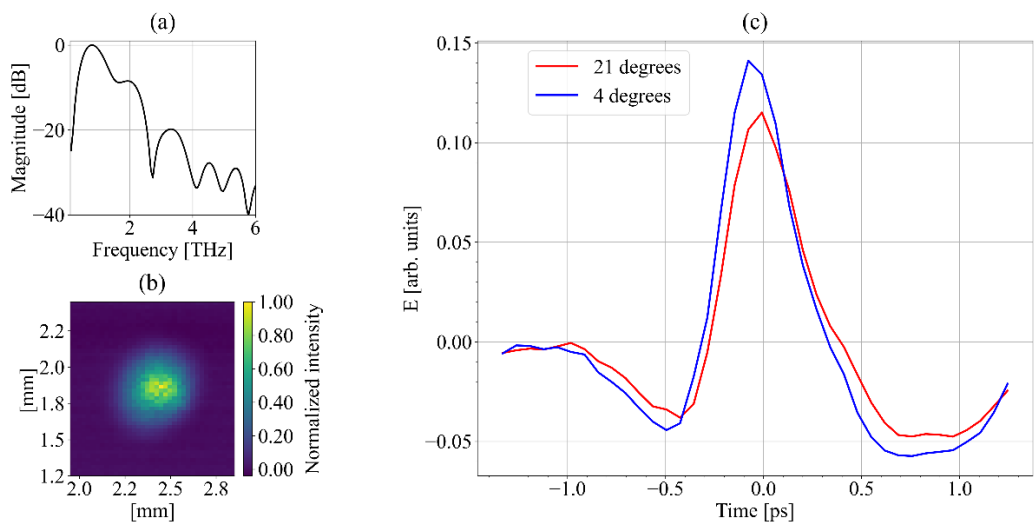

*Figure 1. Terahertz transmission by liquid water in a static cell with diamond windows. a) Example spectrum of the terahertz source. This frequency-dependent spectrum is obtained by Fourier transformation of a typical terahertz field transmitted by an empty path, i.e., without the static liquid cell.*

*b) The spot size of the terahertz beam at the sample position measured with a camera. c) The terahertz fields transmitted by a 100 μm thick layer of liquid water at two temperatures (21 °C and 4 °C, in red and blue, respectively).*

We performed experiments at the beam-line TeraFERMI in Trieste. The source and the available optical setups are detailed elsewhere[30,31]. In short, TeraFERMI generates intense and almost single-cycle THz fields lasting approximately 1 ps with a typical spectrum centered at ~1 THz (Figure 1a). The THz fields are detected via electro-optical sampling[32,33] in a 0.1 mm thick and [110]-oriented gallium phosphide (GaP) crystal with balanced photo-diodes. As the photo-voltage of each single un-balanced diode was not simultaneously detected, it was not possible to estimate the THz field amplitude from the electro-optical coefficients[34]. Due to the finite thickness of the detection crystal, reflections of the sampling and terahertz beams peak at electro-optical delays of about -2 ps and +2 ps, respectively. In order to avoid the spurious contribution of these reflections to the THz spectrum, we obtained Figure 1a by detecting the THz field only between about -1 ps and +1 ps. This windowing induces the apparent frequency oscillations in the spectra of Figure 1a, with a period of about 1 THz. We will use only the value of the THz peak field amplitude in the following analysis. Please note that the spectrum in Figure 1a is just an example of the typical source spectrum. It has been obtained under slightly different experimental conditions and cannot be used as a quantitative reference for the THz fields transmitted by water and shown in Figure 1c. A 2" off-axis parabolic mirror with an effective focal length of 3" focuses the THz field into a static cell with 0.5 mm thick diamond windows enclosing a 100 μm thick layer of pure water. The static cell is magnetically attached to a copper plate, whose temperature is stabilized to either 21 °C or 4 °C by a recirculating chiller, with an accuracy of ±0.1 °C. As shown in Figure 1b, the area of the THz spot size at the sample position was measured to be about 0.6 mm$^2$ full width at half maximum with a camera (Pyrocam IIIHR). A maximum THz power of approximately 390 μW was measured with a pyro-electric based detector (Gentec THZ12D-3S-VP-INT-D0). The intensity lost by reflection at the first air/diamond interface, $r_{ad}^2$, can be estimated to $r_{ad}^2=((1-2.4)/(1+2.4))^2 \sim 17\%$, where $r_{ad}$ is the Fresnel coefficient, 1 is the index of refraction of air, and ~2.4 is the index of refraction of the diamond window[35,36]. Please note that the index of refraction of diamond is practically constant across the frequency[35] (~1 THz) and temperature[36] (4 and 21 °C) ranges investigated here. The intensity lost by reflection at the second interface between diamond and liquid water is approximately $r_{dw}^2=((2.4-2)/(2.4+2))^2 \sim 0.8\%$, where we took the index of refraction of liquid water equal to[37,38] 2, for simplicity. Thus, the THz intensity transmitted by the input diamond window is $(1-r_{ad}^2)(1-r_{dw}^2) \sim 82\%$, considering both air/diamond and diamond/water interfaces. As the maximum THz power available was about 390 μW, we estimate the power transmitted by the front diamond window of the static cell and hitting liquid water to 390μW·82%~320 μW. This results in a maximum THz energy of 320μW/50Hz~6.4 μJ/pulse, fluence of 6.3μJ/0.6mm$^2$~1 mJ/cm$^2$, peak power of 1 mJ/cm$^2$/1ps~1 GW/cm$^2$, and peak field amplitude[39] of $(2 \cdot 376.7\Omega \cdot 1GW/cm^2)^{0.5}$~0.9 MV/cm. Please note that in the estimation of the peak power and field amplitude we assumed, for simplicity, that the temporal shape of the THz pulse is a square with a width of approximately 1 ps. We controlled the THz intensity at the sample with two polarizers with a 10$^4$:1 contrast ratio (infraspecs P01). Finally, we would like to point out that the non-linear response by the diamond windows ($\chi^3$~10$^{-17}$ cm$^2$/V$^2$, see Table 4.1.2 in ref.[27]) is about four orders of magnitude smaller than in pure water ($\chi^3$~10$^{-13}$ cm$^2$/V$^2$, see Table 1 in ref.[26]). Thus, to a first approximation, here we neglect the non-linear contributions by the diamond windows.

The precise estimation of thermal phenomena requires solving intricate differential equations[40]. Here, for simplicity, we assumed that all the THz radiation is absorbed in a uniform layer of water as thick as one penetration depth, and that the temperature decays by thermal diffusion. By using this approach, we were able to estimate the correct order of magnitudes of the thermalization timescales observed in previous experiments[26,38]. Thus, the thermalization time of the sample can be estimated to[38] $\tau \sim L^2/D \sim 17$ ms, where L=50 µm is penetration depth of liquid water at 1 THz, and D=1.5·10$^{-7}$ m$^2$/s is the thermal diffusivity. TeraFERMI emits radiation at a repetition rate of 50 Hz and the time delay between subsequent THz pulses is 20 ms. The thermal relaxation time ($\tau \sim 17$ ms) is comparable to the time-delay between two subsequent pulses (20 ms) and we expect that the temperature build-up of the sample due to the THz pulse train is negligible. Thus, we can over-estimate the maximum temperature increase of the sample due to the absorption of THz radiation from the energy of 2x subsequent pulses[26]. For simplicity, if we assume that two THz pulses are fully and instantaneously absorbed, and transformed into heat in a water layer as thick as one penetration depth at 1 THz (L=50 µm), we over-estimate the maximum temperature increase of the water sample to 2·1mJ/cm$^2$/50µm/(4.2J/(cm$^3$·°C))~0.1 °C. As this is comparable to the accuracy of the recirculating chiller that stabilizes the sample within ±0.1 °C, we can neglect heating effects.

This estimation of the water temperature ignores the heat dissipated onto both diamond windows of the static cell, which reduces the heating of water even further[26]. Previously, we demonstrated that both non-linear refraction and absorption of liquid water at ~1 THz could be estimated correctly from a static liquid cell[24]. The results from water in the static sample holder were identical to the ones obtained by others in a free flowing liquid jet[20,21], within the error bars. Also in that case it was found that heating effects are irrelevant for liquid water at the 50 Hz repetition rate of TeraFERMI[24].

At equilibrium, the linear absorption coefficient of liquid water at THz frequencies depends strongly on temperature[37,38,41,42]. Between about 0.1 and 3 THz, the linear absorption increases upon heating and decreases upon cooling. Based on previous works[37,38,41,42], by decreasing the liquid temperature from 21 °C down to 4 °C, the absorption coefficient of liquid water at 1 THz is expected to drop by $\Delta\alpha_0$=(50±9) cm$^1$. In Figure 1, we show the THz spectrum (Figure 1a), the THz spot size (Figure 1b), and the THz fields transmitted by the diamond sample cell when filled with a 100 µm layer of water at two temperatures, 21 °C and 4°C (Figure 1c). When we cool the liquid to 4 °C, the transmission increases, compared to that at 21 °C, as shown in the blue curve in Figure 1c.

At the electro-optical delay time at which the THz field reaches its maximum value (~0 ps in Figure 1c), all the frequency components of the THz beam add up constructively. Thus, the peak transmission of the THz field reveals the response of the sample that is averaged over the THz spectrum, which peaks at ~1 THz as shown in Figure 1a. The advantage of using the THz peak field is that it can be detected quickly, with less time waited for the movement of the electro-optical sampling delay stage, thus allowing the signal-to-noise ratio to be improved. For these reasons, in the following we discuss only the transmission of the THz peak field. In order to ensure the detection of the THz peak, we always measured a few points around time zero in Figure 1c, i.e., over a temporal range of at least 0.5 ps with 0.07 ps long temporal steps of the electro-optical sampling delay. In this way, we were always able to pinpoint and measure the value of the THz peak. Please note that resolving the full THz pulse is slow here because of the low repetition rate of the source used (50 Hz). In future experiment, we plan to compare these results with the ones obtained with laser-based high repetition rate systems that allow accurate frequency-resolved measurements. For these future experiments, we also plan to measure the complete fluence-dependent response, thus disentangling the absolute value of the nonlinear contribution to the THz transmission.

The ratio between the peak field transmitted by water at 21 °C and the peak field transmitted by water at 4 °C (TR) reveals the temperature-dependent absorption coefficient of liquid water. At equilibrium conditions (linear optics), the ratio between the transmitted THz peak fields at 21 °C and 4 °C is

$$TR = e^{-\Delta\alpha_0 \cdot d/2} \tag{1}$$

where $\Delta\alpha_0 = \alpha_0(21\,°C) - \alpha_0(4\,°C)$ is the difference between the equilibrium absorption coefficients, d is the sample thickness, and the factor 2 accounts for the proportionality between the intensity and electric field squared. Experimentally, when the THz intensity (peak field) impinging upon the water sample is set to 0.3 GW/cm$^2$ (0.5 MV/cm), we find that the ratio between the peak field transmitted by water at 21 °C and the one transmitted at 4 °C is TR=(79.4±0.5)%. By inserting TR=(79.4±0.5)% and d=100 μm in eq.1, we estimate an absorption change of $\Delta\alpha_0$=(46±1.5) cm$^{-1}$ upon cooling water from 21 °C to 4 °C. This value agrees with the literature result at equilibrium[37,38,41,42], $\Delta\alpha_0$=(50±9) cm$^{-1}$. Please note that THz fields with amplitudes of ~0.5 MV/cm do trigger sizeable non-linear signals from water[20–22,24,26]. Thus, this experimental observation implies that the non-linear response of water depends weakly on temperature. In order to investigate this further, we performed additional measurements at increasing THz fields as detailed in the next paragraphs.

When the intensity of the input radiation source is high enough, the absorption coefficient of any material – including liquid water – displays a non-linear response. At simplest, the absorption coefficient becomes intensity-dependent via the equation

$$\alpha(I) = \alpha_0 + \alpha_{NL} \cdot I \tag{2}$$

where $\alpha_0$ is the absorption coefficient of the material at equilibrium introduced previously, I is the source intensity, and $\alpha_{NL}$ is the non-linear term. It has been demonstrated before[20,24,43] that liquid water at room temperature displays non-linear absorption coefficient at ~1 THz equal to $\alpha_{NL}$~-80 cm/GW. As shown in Figure 2, the same value of the transmission ratio was detected at increasing THz intensities (I=0.3, 0.7, 1 GW/cm$^2$) when the liquid sample temperature was kept at °4 C. This experimental finding indicates that the non-linear transmission and absorption by liquid water at 21 °C and 4 °C is the same, within the experimental uncertainty that amounts to about ±0.5% in the transmission ratio or ±1.5 cm$^{-1}$ in absorption.

In other words, we obtain the non-linear and intensity-dependent transmission ratio of a sample, TR(I), by combining eq.1 and eq.2

$$TR(I) = e^{-\Delta\alpha_0 \cdot d/2} e^{-\Delta\alpha_{NL} \cdot I \cdot d/2} \tag{3}$$

where $\Delta\alpha_{NL} = \alpha_{NL}(21\,°C) - \alpha_{NL}(4\,°C)$ is the difference between the non-linear THz absorption coefficients of water at the two temperatures. The transmission ratio described by eq.3 becomes independent of the THz intensity when the argument of the second exponential is negligible, i.e., when $\Delta\alpha_{NL} \cdot I \cdot d/2 \sim 0$. As demonstrated before[20,21,25], the THz intensities used here are enough to induce large non-linear responses in water. Thus, the only condition by which eq.3 becomes independent of the THz intensity is that $\Delta\alpha_{NL}\sim 0$, implying $\alpha_{NL}(21\,°C) \sim \alpha_{NL}(4\,°C)$ within the noise level of this experiment, ~1.5 cm$^{-1}$. Only when this is valid, it is possible to obtain a value of the TR that is independent of the THz intensity, as experimentally demonstrated in Figure 2. In order to estimate the sensibility of these measurements we assume, for simplicity, that the non-linear absorption coefficient of liquid water is zero at 4 °C, $\alpha_{NL}(4\,°C)\sim 0$, and equal to the literature value at[20,24] 21 °C, $\alpha_{NL}$(21 °C)~-80 cm/GW. In this case, the transmission ratio would vary a lot across the intensity range explored here: TR(I=0.3 GW/cm$^2$)~+89% for the minimum THz intensity used, 0.3 GW/cm$^2$, and TR(I=1GW/cm$^2$)~+118% for the maximum one, 1 GW/cm$^2$. However, as shown in Figure 2, we experimentally detect a "flat" TR(I)~79.5% versus intensity, within the experimental noise. Please note that the quantification of the change of the non-linear absorption versus temperature from the theory developed in ref.[20,21] is made difficult by the modified Kramers-Kronig transformation of complex meromorphic functions[44,45].

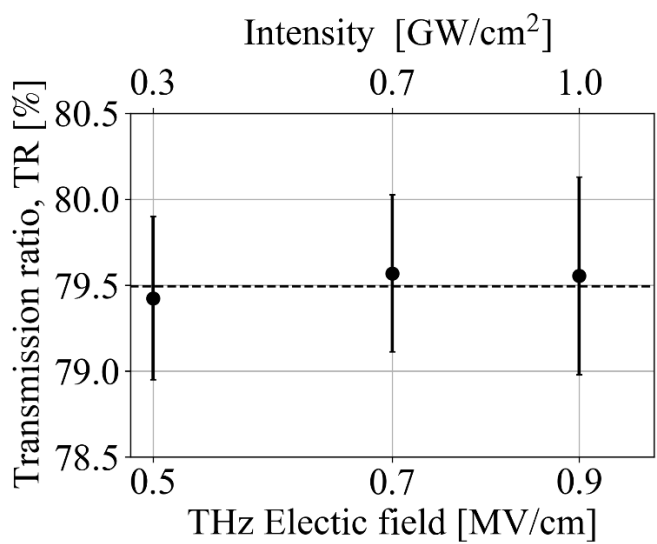

Figure 2. The transmission ratio (TR) is calculated from the THz peak fields transmitted by liquid water at 21 °C and at 4 °C. The TR is evaluated at three intensities (I=0.3, 0.7, and 1 GW/cm$^2$; top axis) or, equivalently, for the maximum terahertz fields shown on the bottom axis (0.5, 0.7, and 0.9 MV/cm). TR measured are the same within the error bars, which are lower than about ±0.5%. Thus, the water transmission at 21 °C and 4 °C is independent on the terahertz intensities probed here. The flat dashed line is a guide to the eye. The error bars are standard errors of the mean from 27 independent measurements.

In conclusion, we measured the THz transmission of a 100-μm thick water layer at two temperatures (21 and 4 °C) as a function of the intensity of the input THz radiation (0.3, 0.7, and 1 GW/cm$^2$). For THz fields high enough to trigger a sizeable non-linearity in water, we showed that the transmission is the same at 21 °C and 4 °C. This implies that the non-linear absorption of liquid water at ~1 THz is equal at these temperatures, within the experimental error (±0.5% in TR and ±1.5 cm$^{-1}$ in absorption).

While several models have been proposed to explain the non-linear THz signals in water[21,22,25,26], the experimental finding reported here suggests that we might prefer models predicting a weakly temperature-dependent non-linear THz response by liquid water. In particular, it is known that both the absorption of the librational band[37] and the generation of solvated electrons[46] depend weakly on temperature. Thus, it is possible that the large non-linear responses detected in water at THz frequencies are related either to tunneling ionization[22,23] or to the resonant excitation of librational motions[25,26]. However, as discussed before[26], tunneling ionization cannot account for the non-linear response of water at 12.3 THz because the ponderomotive energy is about 1000x smaller at 12.3 THz than at 1 THz. For these reasons, the THz fields could be resonantly reorienting hydrogen-bonded water molecules in the liquid phase[25]. Among the different mechanisms proposed to date[21,22,25,26], this dynamic process of THz-driven reorientation[25] might be the most compatible with all of the available experimental results[20–26].

**Acknowledgements**

We acknowledge financial support from the Cluster of Excellence RESOLV (EXC 2033 – 390677874) funded by the Deutsche Forschungsgemeinschaft (DFG, German Research Foundation) and by the ERC Advanced Grant 695437 (THz Calorimetry). F.N. acknowledges funding by the DFG with project 509442914. We acknowledge support by the DFG Open Access Publication Funds of the Ruhr-Universität Bochum. These results are part of a project that has received funding from the European Research Council (ERC) under the European Union's Horizon 2020 research and innovation programme (grant agreement No. 805202 - Project Teraqua). This project received funding from the European Union's Horizon 2020 research and innovation programme under the Marie Skłodowska-Curie grant agreement No 801459 - FP-RESOMUS. We thank CNR-IOM for the use of the MENLO C-Fiber780 laser. We are grateful to P. Di Pietro, A. Perucchi, and M. Havenith for support and discussions.

**Conflict of Interest**

There are no conflicts to declare.

**Data availability**

The data that supports the findings of this study are available within the article.

**References**

[1] D. Laage and J.T. Hynes, Science **311**, 832 (2006).

[2] A. Nilsson and L.G.M. Pettersson, Nat. Commun. **6**, 8998 (2015).

[3] P. Ball, Nature **452**, 291 (2008).

[4] J.K. Vij, D.R.J. Simpson, and O.E. Panarina, J. Mol. Liq. **112**, 125 (2004).

[5] F. Novelli, B. Guchhait, and M. Havenith, Materials (Basel). **13**, 1311 (2020).

[6] M. Cho, G.R. Fleming, S. Saito, I. Ohmine, and R.M. Stratt, J. Chem. Phys. **100**, 6672 (1994).

[7] M. Heyden, J. Sun, S. Funkner, G. Mathias, H. Forbert, M. Havenith, and D. Marx, Proc. Natl. Acad. Sci. **107**, 12068 (2010).

[8] D.C. Elton and M. Fernández-Serra, Nat. Commun. **7**, 10193 (2016).

[9] I. Popov, P. Ben Ishai, A. Khamzin, and Y. Feldman, Phys. Chem. Chem. Phys. **18**, 13941 (2016).

[10] D.C. Elton, Phys. Chem. Chem. Phys. **19**, 18739 (2017).

[11] C. Hölzl, H. Forbert, and D. Marx, Phys. Chem. Chem. Phys. **23**, 20875 (2021).

[12] S. Ahmed, A. Pasti, R.J. Fernández-Terán, G. Ciardi, A. Shalit, and P. Hamm, J. Chem. Phys. **148**, 234505 (2018).

[13] J. Savolainen, S. Ahmed, and P. Hamm, Proc. Natl. Acad. Sci. **110**, 20402 (2013).

[14] A. Shalit, S. Ahmed, J. Savolainen, and P. Hamm, Nat. Chem. **9**, 273 (2017).

[15] A. Berger, G. Ciardi, D. Sidler, P. Hamm, and A. Shalit, Proc. Natl. Acad. Sci. U. S. A. **116**, 2458 (2019).

[16] M.C. Hoffmann, N.C. Brandt, H.Y. Hwang, K.-L. Yeh, and K.A. Nelson, Appl. Phys. Lett. **95**, 231105 (2009).

[17] P. Zalden, L. Song, X. Wu, H. Huang, F. Ahr, O.D. Mücke, J. Reichert, M. Thorwart, P.K. Mishra, R. Welsch, R. Santra, F.X. Kärtner, and C. Bressler, Nat. Commun. **9**, 2142 (2018).

[18] H. Elgabarty, T. Kampfrath, D.J. Bonthuis, V. Balos, N.K. Kaliannan, P. Loche, R.R. Netz, M. Wolf, T.D. Kühne, and M. Sajadi, Sci. Adv. **6**, 1 (2020).

[19] V. Balos, N.K. Kaliannan, H. Elgabarty, M. Wolf, T.D. Kühne, and M. Sajadi, Nat. Chem. **14**, 1031 (2022).

[20] M.O. Zhukova, M. V. Melnik, A.N. Tcypkin, I.O. Vorontsova, S.E. Putilin, S.A. Kozlov, X.-C.C. Zhang, M. V. Melnik, M.O. Zhukova, I.O. Vorontsova, S.E. Putilin, S.A. Kozlov, and X.-C.C. Zhang, Opt. Express **27**, 10419 (2019).

[21] A. Tcypkin, M. Zhukova, M. Melnik, I. Vorontsova, M. Kulya, S. Putilin, S. Kozlov, S. Choudhary, and R.W. Boyd, Phys. Rev. Appl. **15**, 054009 (2021).

[22] A. Ghalgaoui, L.-M. Koll, B. Schütte, B.P. Fingerhut, K. Reimann, M. Woerner, and T. Elsaesser, J. Phys. Chem. Lett. **11**, 7717 (2020).

[23] A. Ghalgaoui, B.P. Fingerhut, K. Reimann, T. Elsaesser, and M. Woerner, Phys. Rev. Lett. **126**, 097401 (2021).

[24] F. Novelli, C.Y. Ma, N. Adhlakha, E.M. Adams, T. Ockelmann, D. Das Mahanta, P. Di Pietro, A. Perucchi, and M. Havenith, Appl. Sci. **10**, 5290 (2020).

[25] F. Novelli, L. Ruiz Pestana, K.C. Bennett, F. Sebastiani, E.M. Adams, N. Stavrias, T. Ockelmann, A. Colchero, C. Hoberg, G. Schwaab, T. Head-Gordon, and M. Havenith, J. Phys. Chem. B **124**, 4989 (2020).

[26] F. Novelli, C. Hoberg, E.M. Adams, J.M. Klopf, and M. Havenith, Phys. Chem. Chem. Phys. **24**, 653 (2022).

[27] R.W. Boyd, *Nonlinear Optics*, 3rd ed. (Academic Press, 2007).

[28] J.E. Bertie and Z. Lan, Appl. Spectrosc. **50**, 1047 (1996).

[29] J. Prakash, M.M. Seyedebrahimi, A. Ghazaryan, J. Malekzadeh-Najafabadi, V. Gujrati, and V. Ntziachristos, Proc. Natl. Acad. Sci. **117**, 4007 (2020).

[30] P. Di Pietro, N. Adhlakha, F. Piccirilli, L. Capasso, C. Svetina, S. Di Mitri, M. Veronese, F. Giorgianni, S. Lupi, and A. Perucchi, Synchrotron Radiat. News **30**, 36 (2017).

[31] U. Happek, A.J. Sievers, and E.B. Blum, Phys. Rev. Lett. **67**, 2962 (1991).

[32] P.C.M. Planken, H.-K. Nienhuys, H.J. Bakker, and T. Wenckebach, J. Opt. Soc. Am. B **18**, 313 (2001).

[33] S. Casalbuoni, H. Schlarb, B. Schmidt, P. Schmüser, B. Steffen, and A. Winter, Phys. Rev. Spec. Top. - Accel. Beams **11**, 072802 (2008).

[34] H. Hirori, A. Doi, F. Blanchard, and K. Tanaka, Appl. Phys. Lett. **98**, 091106 (2011).

[35] E. Bründermann, H.-W. Hübers, and M.F. Kimmitt, *Terahertz Techniques* (Springer Berlin Heidelberg, Berlin, Heidelberg, 2012).

[36] V.Y. Yurov, E. V. Bushuev, A.F. Popovich, A.P. Bolshakov, E.E. Ashkinazi, and V.G. Ralchenko, J. Appl. Phys. **122**, 243106 (2017).

[37] H.R. Zelsmann, J. Mol. Struct. **350**, 95 (1995).

[38] F. Novelli, J.W.M. Chon, and J.A. Davis, Opt. Lett. **41**, 5801 (2016).

[39] B. Liu, H. Bromberger, A. Cartella, T. Gebert, M. Först, and A. Cavalleri, Opt. Lett. **42**, 129 (2017).

[40] A.D. Kraus and A. Bejan, *Heat Transfer Handbook* (John Wiley & Sons, Inc., Hoboken, New Jersey, 2003).

[41] P. Lunkenheimer, S. Emmert, R. Gulich, M. Köhler, M. Wolf, M. Schwab, and A. Loidl, Phys. Rev. E **96**, 062607 (2017).

[42] W.J. Ellison, J. Phys. Chem. Ref. Data **36**, 1 (2007).

[43] X. Zheng, R. Chen, G. Shi, J. Zhang, Z. Xu, X. Cheng, and T. Jiang, Opt. Lett. **40**, 3480 (2015).

[44] K.-E. Peiponen, V. Lucarini, J.J. Saarinen, and E. Vartiainen, Appl. Spectrosc. **58**, 499 (2004).

[45] K.-E. Peiponen, J. Phys. A. Math. Gen. **34**, 6525 (2001).

[46] F.Y. Jou and G.R. Freeman, J. Phys. Chem. **83**, 2383 (1979).